\documentclass[12pt,psfig,preprint]{aastex}

\shorttitle{Dark energy and the angular size - redshift diagram}
\shortauthors{Lima \& Alcaniz}

\begin{document}

\title{Dark energy and the angular size - redshift diagram for milliarcsecond
radio-sources}

\author{J. A. S. Lima\altaffilmark{1} and J. S. Alcaniz\altaffilmark{2}}
\affil{Departamento de F\'{\i}sica, Universidade Federal do Rio Grande do 
Norte, 
\\ 
C.P. 1641, 59072-970, Natal, Brasil}

\altaffiltext{1}{limajas@dfte.ufrn.br}
\altaffiltext{2}{alcaniz@dfte.ufrn.br}

\begin{abstract}
We investigate observational constraints on the cosmic equation
of state  from measurements of angular size for a large sample of
milliarcsecond compact radio-sources. The results are based on a flat
Friedmann-Robertson-Walker (FRW) type models driven by non-relativistic matter
plus a smooth dark energy component parametrized by its equation of state $p_x
= \omega \rho_x$ ($-1 \leq \omega < 0$). The allowed intervals for $\omega$ and
$\Omega_{\rm{m}}$ are heavily dependent on the value of the mean projected
linear size $l$. For $l \simeq  20h^{-1}  - 30h^{-1}$ pc, we find
$\Omega_{\rm{m}} \leq 0.62$, $\omega \leq -0.2$ and $\Omega_{\rm{m}} \leq
0.17$, $\omega \leq -0.65$ (68$\%$ c.l.), respectively. As a general
result, this analysis shows that if one minimizes $\chi^{2}$ for the
parameters $l$, $\Omega_{\rm{m}}$ and $\omega$, the conventional
flat $\Lambda$CDM model ($\omega = -1$) with  $\Omega_{\rm{m}} = 0.2$ and $l
= 22.6 h^{-1}$pc is the best fit for these angular size data.
\end{abstract}


A large number of recent observational evidences strongly suggest that we live
in  a flat, accelerating Universe composed by $\sim$ 1/3 of matter (barionic
+ dark) and  $\sim$ 2/3 of an exotic component with large negative pressure,
usually named dark energy or ``quintessence". The basic set of experiments
includes: observations from SNe Ia (Perlmutter {\it et al.} 1998; 1999; Riess
et al. 1998), CMB anisotropies (de Bernardis {\it et al.} 2000), large scale
structure (Bahcall 2000), age estimates of globular clusters (Carretta {\it et
al.} 2000; Krauss 2000; Rengel {\it et al.} 20001) and old high redshift
galaxies (OHRG's) (Dunlop 1996; Krauss 1997; Alcaniz \& Lima 1999; Alcaniz \&
Lima 2001). It is now believed that such results provide the remaining piece
of information connecting the inflationary flatness prediction
($\Omega_{\rm{T}} = 1$) with astronomical observations, and, perhaps more
important from a theoretical viewpoint, they have stimulated the current
interest for more general models containing an extra component describing this
unknown dark energy, and simultaneously accounting for the present accelerated
stage of the Universe.  

The absence of a convincing evidence concerning the nature of
this dark component gave origin to an intense debate and mainly to many
theoretical  speculations in the last few years. Some possible candidates for
``quintessence" are: a vacuum decaying energy density, or a time varying
$\Lambda$-term (Ozer \& Taha 1987; Freese 1987; Carvalho {\it et al} 1992,
Lima and Maia 1994), a relic scalar field (Peebles \& Ratra 1988; Frieman
{\it et al} 1995; Caldwell {\it et al} 1998; Saini {\it et al} 2000) or still
an extra component, the so-called ``X-matter", which is simply characterized
by an equation of state $p_x=\omega\rho_{x}$, where $\omega \geq -1$ (Turner
\& White 1997; Chiba {\it et al} 1997) and includes, as a particular case,
models with a cosmological constant ($\Lambda$CDM) (Peebles 1984). For
``X-matter" models, several results suggest $\Omega_x \simeq 0.7$ and $\omega
\leq -0.6$. For example, studies from gravitational lensing + SNe Ia provide
$\omega \leq -0.7$ at 68$\%$ c.l. (Waga \& Miceli 1999; see also Dev {\it et al.}
2001). Limits from age estimates of old galaxies at high redshifts require 
$\omega < -0.27$ for $\Omega_{\rm{m}} \simeq 0.3$ (Lima  \& Alcaniz 2000a).
In addition, constraints from large scale structure (LSS) and cosmic microwave
background anisotropies (CMB) complemented by the SN Ia data, require $0.6
\leq \Omega_x \leq 0.7$ and $\omega < -0.6$ ($95\%$ c.l.) for a flat universe
(Garnavich {\it et al} 1998; Perlmutter {\it et al} 1999; Efstathiou 1999),
while for universes with arbitrary spatial curvature these data provide
$\omega < -0.4$ (Efstathiou 1999). 

On the other hand, although carefully investigated in many of their
theoretical and  observational aspects, an overview on the literature shows
that a quantitative analysis on the influence of a ``quintessence" component 
($\omega ={p_x}/{\rho_x}$) in some kinematic tests like angular size-redshift
relation still remains to be analysed.  Recently, Lima \& Alcaniz (2000b)
studied some qualitative aspects of this test in the context of such models,
with particular emphasis for the critical redshift $z_m$ at which the angular
size takes its minimal value. As a general conclusion, it was shown that this
critical redshift cannot discriminate between world models since different
scenarios can provide similar values of $z_m$ (see also Krauss \& Schramm
1993). This situation is not improved even if evolutionary effects are taken
into account. In particular, for the observationally favoured open universe
($\Omega_{\rm{m}} = 0.3$) we found $z_m=1.89$, a value that can also be
obtained for quintessence models having  $0.85 \leq \Omega_x \leq 0.93$ and
$-1 \leq \omega_x \leq -0.5$. Qualitatively, it was also argued that if the
predicted $z_m$ is combinated with other tests, some interesting cosmological
constraints can be obtained.

In this letter, we focus our attention on a more quantitative analysis. We
consider the $\theta - z$ data of compact radio sources recently updated and
extended by Gurvits {\it et al.} (1999) to constrain the cosmic equation of
state. We show that a good agreement between theory and observation is
possible if $\Omega_{\rm{m}} \leq 0.62$, $\omega \leq -0.2$ and
$\Omega_{\rm{m}} \leq 0.17$, $\omega \leq -0.65$ (68$\%$ c.l.) for values of
the mean projected linear size between $l \simeq  20h^{-1}  - 30h^{-1}$ pc,
respectively. In particular we find that a conventional cosmological constant
model ($\omega = -1$) with $\Omega_{\rm{m}} = 0.2$ and  $l = 22.64h^{-1}$ pc
is the best fit model for these data with $\chi^{2} = 4.51$ for 9 degrees of
freedom.

For spatially flat, homogeneous,
and isotropic cosmologies driven by nonrelativistic matter plus an exotic
component with equation of state, $p_{x} = \omega\rho_{x}$, the Einstein field
equations  take the following form:   
\begin{equation} 
({\dot{R} \over R})^{2} = H_{o}^{2}\left[\Omega_{\rm{m}}({R_{o} \over R})^{3} +  
\Omega_x({R_{o} \over R})^{3(1 + \omega)}\right] , 
\end{equation} 
\begin{equation} 
{\ddot{R} \over R} = -{1 \over 2}H_{o}^{2}\left[\Omega_{\rm{m}}({R_{o} \over 
R})^{3} +  
(3\omega + 1)\Omega_x({R_{o} \over R})^{3(1 + \omega)}\right]  , 
\end{equation} 
where an overdot denotes derivative with respect to time, $H_{o} = 100h
{\rm{Kms^{-1}Mpc^{-1}}}$  is the   present value of the Hubble parameter, and 
$\Omega_{\rm{m}}$ and $\Omega_x$ are the present day  matter  and quintessence
density parameters. As one may check  from (1) and (2), the case $\omega= - 1$
corresponds effectively to a cosmological constant.

In such a background, the angular size-redshift relation for a rod of 
intrinsic length $l$ can be written as (Sandage 1988) 
\begin{equation} 
\theta(z) = {D (1 + z) \over \xi(z)} \quad.
\end{equation}
In the above expression $D$ is the angular-size scale expressed in
milliarcseconds (marcs)
\begin{equation}
D = {100 lh \over c} ,
\end{equation}
where $l$ is measured in parsecs (for compact radio-sources), and the
dimensionless coordinate  $\xi$ is given by (Lima \& Alcaniz 2000b) 
\begin{equation}  
\xi(z) = \int_{(1 + z)^{-1}}^{1} {dx \over x \left[\Omega_{\rm{m}}x^{-1} +
(1 - \Omega_{\rm{m}}) x^{-(1 + 3\omega)}\right]^{{1}\over{2}}}. 
\end{equation} 

The above equations imply that for given values of $l$,
$\Omega_{\rm{m}}$ and $\omega$, the predicted value of $\theta(z)$  is
completely determined. Two points, however, should be stressed before
discussing the resulting diagrams. First of all, the determination of $\Omega_{\rm{m}}$
and $\omega$ are strongly dependent on the adopted value of $l$.
In this case, instead of assuming a especific value for the mean projected
linear size,  we have worked on the interval $l \simeq 20h^{-1}  - 30h^{-1}$
pc, i.e., $l \sim O(40)$ pc for $h = 0.65$, or equivalently, $D = 1.4 - 2.0$
marcs. Second, following Kellermann (1993), we assume that possible
evolutionary effects can be removed out from this sample because compact radio
jets are (i) typically  less than a hundred parsecs in extent, and, therefore,
their morphology and kinematics do not depend considerably on the
intergalactic medium and (ii) they have typical ages of some tens of years,
i.e., they are very young compared to the age of the Universe.

In our analysis we consider the angular size data for
milliarcsecond radio-sources recently compiled by Gurvits {\it et al.} (1999). This
data set, originally composed by 330 sources distributed over a wide range of redshifts
($0.011 \leq z \leq 4.72$), was reduced to 145 sources with spectral index 
$-0.38 \leq \alpha \leq 0.18$ and total luminosity $Lh^{2} \geq 10^{26}$ W/Hz
in order  to minimize any possible dependence of angular size on spectral index
and/or linear size on  luminosity. This  new sample was  distributed into 12
bins with  12-13 sources per bin (see Fig. 1). In order to determine the
cosmological parameters $\Omega_{\rm{m}}$ and $\omega$,  we use a $\chi^{2}$
minimization neglecting the unphysical region $\Omega_{\rm{m}} < 0$, 
\begin{equation}    
\chi^{2}(l, \Omega_{\rm{m}}, \omega) =
\sum_{i=1}^{12}{\frac{\left[\theta(z_{i}, l, \Omega_{\rm{m}}, \omega) -
\theta_{oi}\right]^{2}}{\sigma_{i}^{2}}},   
\end{equation}  
where $\theta(z_{i}, l, \Omega_{\rm{m}}, \omega)$ is given by Eqs. (3)
and (5) and $\theta_{oi}$ is the observed values of the angular size
with errors $\sigma_{i}$ of the $i$th bin in  the sample.

Figure 1 displays the binned data of the median angular size plotted against
redshift. The curves represent flat quintessence models with $\Omega_{\rm{m}}
= 0.3$ and some selected values of $\omega$. As discussed in Lima \& Alcaniz
(2000b), the standard open model (thick line) may be interpreted as an
intermediary case between $\Lambda$CDM ($\omega = -1$) and quintessence models
with $\omega \leq -0.5$. In Fig. 2  we show contours of constant likelihood
(95$\%$ and 68$\%$) in the plane  $\omega - \Omega_{\rm{m}}$ for the interval
$l \simeq 20h^{-1}  - 30h^{-1}$ pc.  For $l = 20.58h^{-1}$ pc ($D = 1.4$
marcs),  the best fit occurs for $\Omega_{\rm{m}} = 0.26$ and $\omega =
-0.86$. As can be seen there, this assumption provides $\Omega_{\rm{m}} \leq 
0.48$ and $\omega = -0.3$ at 1$\sigma$. In the subsequent panels of the same
figure similar analyses are  displayed for $l \simeq 22.05h^{-1}$ pc ($D =
1.5$ marcs), $l \simeq 23.53h^{-1}$ pc ($D = 1.6$ marcs) and $l \simeq
29.41h^{-1}$ pc ($D = 2.0$ marcs), respectively.  As should be physically
expected, the limits are now much more restrictive than in the previous case
because for the same values of $\theta_{oi}$ it is needed larger $\xi(z)$ (for
larger $l$) and, therefore, smaller values of $\omega$. For $l \simeq
29.41h^{-1}$ pc, we find $\Omega_{\rm{m}} = 0.04$ and $\omega = -1$ as the
best fit model.  For intermediate values of $l$, namely, $l = 22.0h^{-1}$ pc
($D = 1.5$ marcs) and $l = 23.5h^{-1}$ pc ($D = 1.6$ marcs), we have
$\Omega_{\rm{m}} = 0.22$, $\omega = -0.98$ and $\Omega_{\rm{m}} = 0.16$ and
$\omega = -1$, respectively. In particular, for smaller values of $l$, e.g.,
$l \simeq 14.70h^{-1}$ pc ($D = 1.0$ marcs) we find $\Omega_{\rm{m}} = 0.36$,
$\omega = -0.04$. As a general result (independent of the choice of $l$), if
we minimize $\chi^{2}$ for $l$, $\Omega_{\rm{m}}$ and $\omega$, we obtain   $l
= 22.64h^{-1}$ pc ($D = 1.54$ marcs),  $\Omega_{\rm{m}} = 0.2$  and $\omega =
-1$ with $\chi^{2} = 4.51$ for 9 degrees of freedom (see Table 1). It is worth
notice that our results are rather different from those presented by Jackson
\& Dodgson (1996) based on the original Kellermann's data (Kellermann 1993).
They argued that the Kellermann's compilation favours open and highly
decelerating models with negative cosmological constant. Later on, they
considered a bigger sample of 256 sources selected from the compilation of
Gurvits (1994) and concluded that the standard flat CDM model is ruled out at
$98.5\%$ confidence level whereas low-density models with a cosmological
constant of either sign are favoured (Jackson \& Dodgson 1997). More recently,
Vishwakarma (2001) used  the updated data of Gurvits {\it et al.} (1999) to
compare varying and constant $\Lambda$CDM models. He concluded that flat
$\Lambda$CDM models with $\Omega_{\rm{m}} = 0.2$ are favoured.

At this point it is also interesting to compare our results with some recent
determinations of $\omega$  derived from independent methods. Recently,
Garnavich {\it et al.} (1998) using the SNe Ia data from the High-Z Supernova
Search Team found $\omega < -0.55$  ($95\%$ c.l.) for flat models  whatever
the value of $\Omega_{\rm{m}}$ whereas for arbitrary geometry they obtained
$\omega < -0.6$ ($95\%$ c.l.). As commented there, these values are
inconsistent with an unknown component like topological defects (domain walls,
string, and textures) for which $\omega = -\frac{n}{3}$, being $n$ the
dimension of the defect. The results by Garnavich {\it et al.} (1998) agree
with the constraints obtained from a wide variety of different phenomena 
(Wang {\it et al.} 1999), using the ``concordance cosmic" method. Their
combined maximum likelihood analysis suggests $\omega \leq -0.6$, which is
less stringent than the upper limits derived here for values of $l \geq
20h^{-1}$ pc. More recently, Balbi {\it et al.} (2001) investigated CMB
anisotropies in quintessence models by using the MAXIMA-1 and BOOMERANG-98
published  bandpowers in combination with the COBE/DMR results (see also
Corasaniti \& Copeland 2001). Their analysis sugests $\Omega_x > 0.7$ and $-1
\leq \omega \leq -0.5$ whereas Jain {\it et al} (2001) found,  by using image
separation distribution function of lensed quasars, $-0.75 \leq \omega \leq
-0.42$, for the observed range of $\Omega_m \sim 0.2 - 0.4$ (Dekel {\it et
al.} 1997). These and other recents results are summarized in Table 2.

Let us now discuss briefly these angular  size constraints whether the
adopted X-matter model  is replaced by a scalar field motivated cosmology, as
for instance, that one proposed by Peebles and Ratra (1988). These models are
defined by power law potentials, $V(\phi) \sim \phi^{- \alpha}$, in such a way
that the parameter of the effective equation of state ($w_\phi = p_\phi /
\rho_\phi$) may become constant at late times (or for a given era). In this
case, as shown elsewhere (Lima \& Alcaniz 2000c), the dimensionless quantity
$\xi$ defining the angular size reads 
\begin{equation} 
\xi(z) = \int_{(1 + z)^{-1}}^{1} {dx \over x [\Omega_{\rm{m}}x^{-1} + (1 -
\Omega_{\rm{m}}) x^{{{4-\alpha}  \over {2 + \alpha}}}]^{{1}\over{2}}}. 
\end{equation}   
Comparing the above expression with (5) we see that if $\omega = -{2/(2 +
\alpha)}$ this class of models may reproduce faithfully the X-matter
constraints based on the angular size observations presented here. However, as
happens with the Supernovae type Ia data (Podariu \& Ratra 2000), the two sets
of confidence contours may differ significantly if one goes beyond the time
independent equation of state approximation. Naturally, a similar behavior is
expected if generic scalar field potentials are considered.   

Finally, we stress that measurements of angular size from distant sources
provide an important test for world models. However, in order to improve the
results a  statistical study describing the intrinsic lenght distribution of
the sources seems to be of fundamental importance. On the other hand, in the
absence of such analysis but living in the era of {\it precision cosmology},
one may argue that reasonable values for astrophysical quantities (like the
characteristic linear size $l$) can be infered from the best cosmological
model. As observed by Gurvits (1994), such an estimative could be useful for
any kind of study envolving physical parameters of active galactic nuclei
(AGN). In principle, by  knowing $l$ and assuming a physical model for AGN, 
a new method to estimate the Hubble parameter could be established. 

\section*{Acknowledgments}

The authors are grateful to L. I. Gurvits for sending his compilation of the data as well as for 
helpful discussions. We would like to thank Gang Chen for useful discussions.  This
work was  partially suported by the Conselho Nacional de Desenvolvimento
Cient\'{\i}fico e  Tecnol\'{o}gico - CNPq,  Pronex/FINEP (no. 41.96.0908.00) 
and FAPESP (00/06695-0).


\clearpage

\begin{figure}
\plotone{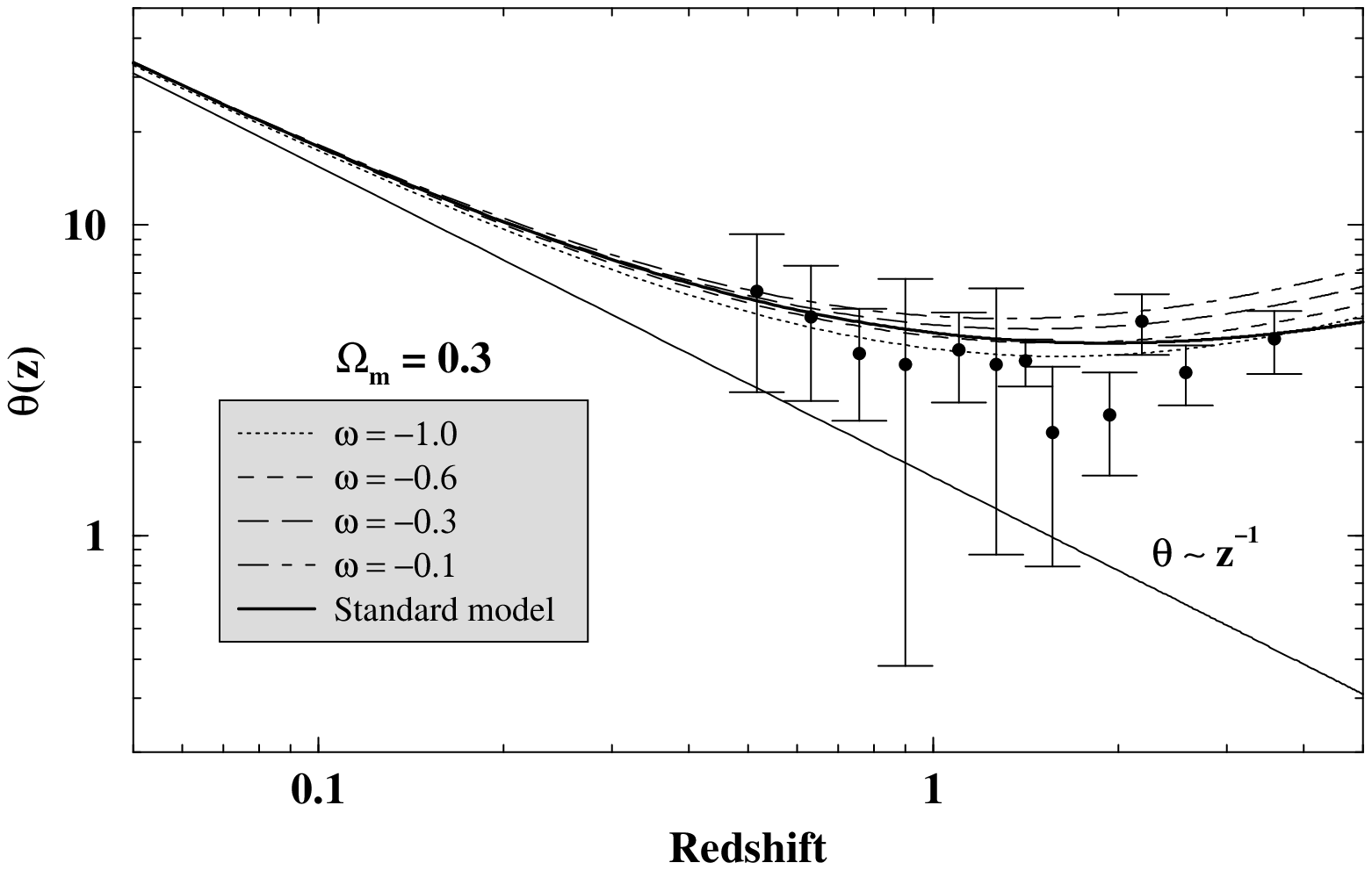}
\caption{Angular size versus redshift for 145 sources binned
into 12 bins (Gurvits {\it et al.} 1999). The curves correspond to the
characteristic linear size $l = 22.64 h^{-1}$ pc. Thick solid curve is the
prediction of the standard open model ($\Omega_{\rm{m}} = 0.3$).}
\end{figure}

\clearpage 

\begin{figure}
\plotone{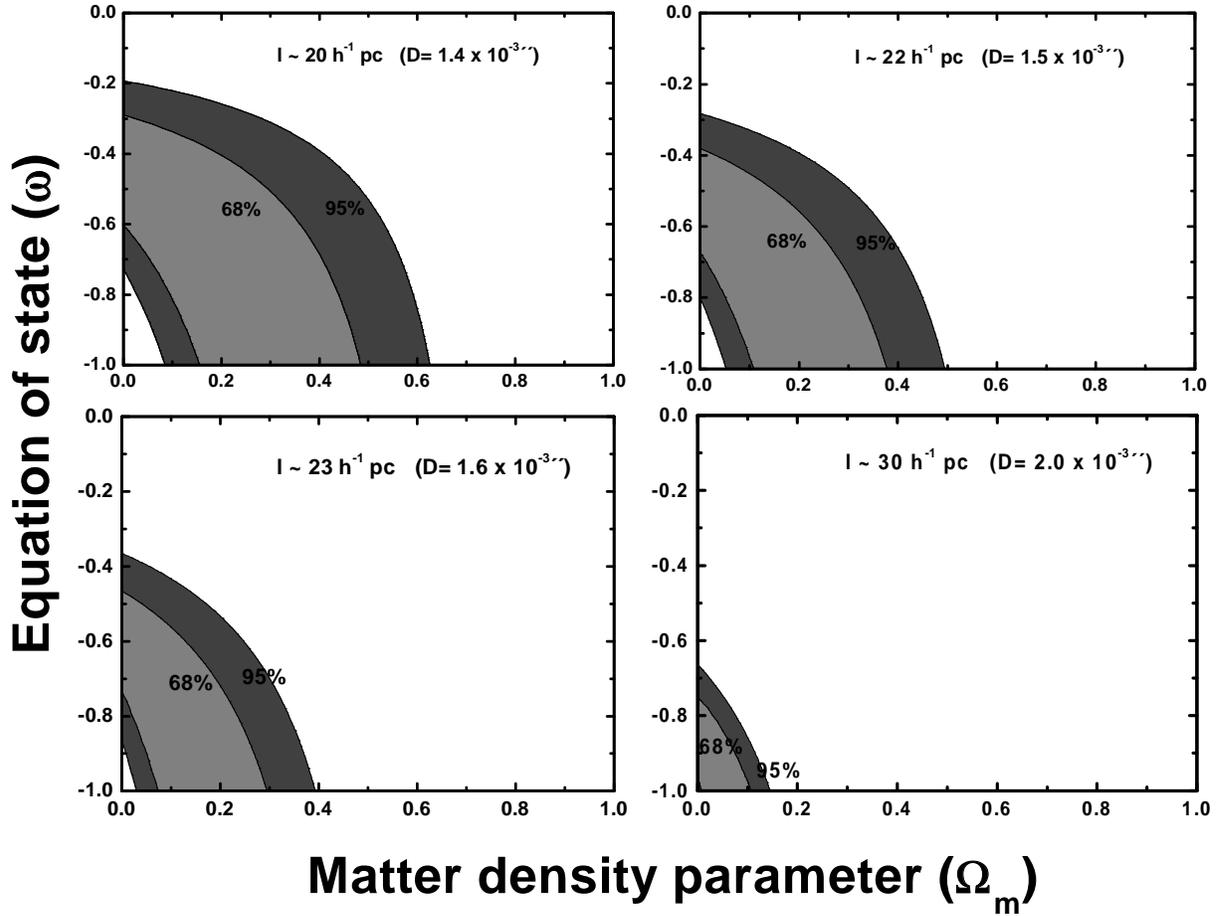}
\caption{Confidence regions in the $\omega - \Omega_m$ plane according to the
updated sample of angular size data (Gurvits {\it et al.} 1999). The solid
lines in each panel show the 95$\%$ and 68$\%$ likelihood
contours for flat quintessence models.}
\end{figure}

\clearpage

\begin{table}  
\begin{center}  
\begin{tabular}{rrrlll}  
\hline  \hline \\
\multicolumn{1}{c}{$D$ (mas)}&
\multicolumn{1}{c}{$lh$ (pc)}& 
\multicolumn{1}{c}{$\Omega_{\rm{m}}$}& 
\multicolumn{1}{c}{$\omega$}& 
\multicolumn{1}{c}{$\chi^{2}$}\\
\\  \hline  \hline
1.4& 20.58& 0.26& -0.86& 4.56\\ 
\\
1.5& 22.05& 0.22& -0.98& 4.52\\
\\
1.6& 23.53& 0.16& -1& 4.54\\ 
\\
2.0& 29.41& 0.04& -1& 5.57\\ 
\\
Best fit: 1.54& 22.64& 0.2& -1& 4.51\\ 
\hline  \hline
\\
\end{tabular} 
\begin{center} 
\caption{Limits on $\omega$ from $\theta - z$ relation}
\end{center}  
\end{center} 
\end{table}

\begin{table} 
\begin{center}  
\begin{tabular}{rlll}
\hline \hline
\\
\multicolumn{1}{c}{Method}& 
\multicolumn{1}{c}{Author}& 
\multicolumn{1}{c}{$\Omega_{\rm{m}}$}& 
\multicolumn{1}{c}{$\omega$}\\ 
\\ 
\hline \hline 
CMB+SNe Ia..& Turner \& White (1997)& $\simeq 0.3$& $\simeq
-0.6$\\   
 & Efstathiou (1999)& $\sim$&  $< -0.6$\\ 
SNe Ia............& Garnavich {\it et al.} (1998)&  
$\sim$& $< -0.55$\\ 
SGL+SNe Ia..& Waga \& Miceli (1999)& 
$0.24$& $< -0.7$\\ 
SNe Ia+LSS...& Perlmutter {\it et al.} (1999)& $\sim$ & $< -0.6$\\ 
Various............& Wang {\it et al.} (1999)& $0.2 - 0.5$& $ < -0.6$\\ 
OHRG`s..........& Lima \& Alcaniz (2000a)&  
$0.3$& $\leq -0.27$\\ 
CMB...............& Balbi {\it et al.} (2001)&  
$0.3$& $\leq -0.5$\\ 
& Corasaniti \& Copeland  (2001)&  
$\sim$& $\leq  -0.96$\\ 
SGL................& Jain {\it et al.} (2001)& $0.2 - 0.4$& $\geq
-0.75$,  $\leq -0.55$\\ 
\hline 
\hline 
\end{tabular} 
\begin{center}  
\caption{Limits to $\omega$ for a given $\Omega_{\rm{m}}$} 
\end{center}  
\end{center}
\end{table}

\end{document}